\documentclass[conference]{IEEEtran}
\IEEEoverridecommandlockouts
\usepackage{cite}
\usepackage{amsmath,amssymb,amsfonts}
\usepackage{algorithmic}
\usepackage{graphicx}
\usepackage{textcomp}
\usepackage{xcolor}
\usepackage{url}
\usepackage{authblk}
\usepackage{multirow}

\def\BibTeX{{\rm B\kern-.05em{\sc i\kern-.025em b}\kern-.08em
    T\kern-.1667em\lower.7ex\hbox{E}\kern-.125emX}}
\begin{document}

\title{VecIntrinBench: Benchmarking Cross-Architecture Intrinsic Code Migration for RISC-V Vector}

\author[1,2]{Liutong Han}
\author[1,2]{Chu Kang}
\author[1]{Mingjie Xing\textsuperscript{\textdagger}\thanks{\textsuperscript{\textdagger}Corresponding author: Mingjie Xing (mingjie@iscas.ac.cn)}}
\author[1]{Yanjun Wu}

\affil[1]{Intelligent Software Research Center, Institute of Software, Chinese Academy of Sciences, Beijing, China}
\affil[2]{University of Chinese Academy of Sciences, Beijing, China}

\maketitle

\begin{abstract}

Intrinsic functions are specialized functions provided by the compiler that efficiently operate on architecture-specific hardware, allowing programmers to write optimized code in a high-level language that fully exploits hardware features. Using Single Instruction, Multiple Data (SIMD) intrinsics to vectorize core code blocks is a standard optimization method in high-performance libraries, often requiring specific vector optimization implementations for multiple mainstream architectures. The promising RISC-V software ecosystem has a significant demand for algorithm library migration and adaptation. Translating existing intrinsic functions to RISC-V Vector (RVV) intrinsic functions across architectures is currently a mainstream approach. Rule-based intrinsic mapping methods and LLM-based code generation can help developers address the code migration challenge. However, existing intrinsic code benchmarks focus on mainstream SIMD intrinsics and lack support for the emerging RISC-V architecture. There is currently no benchmark that comprehensively evaluates the intrinsic migration capabilities for the RVV extension. To fill this gap, we propose VecIntrinBench, the first intrinsic benchmark encompassing RVV extensions. It includes 50 function-level tasks from open source repositories, implemented as scalars, RVV intrinsics, Arm Neon intrinsics, and x86 intrinsics, along with comprehensive functional and performance test cases. We systematically evaluated various code migration approaches on VecIntrinBench, yielding a series of insightful findings. The results demonstrate that advanced Large Language Models (LLMs) achieve a similar effect as rule-based mapping approaches for RISC-V code migration, while also delivering superior performance. We further analyze the reasons and identify future directions for LLM development in the code migration field. The VecIntrinBench is open-sourced to benefit the broader community and developers.

\end{abstract}

\begin{IEEEkeywords}
RISC-V Vector Intrinsic, Code Migration, Benchmark, Large Language Model
\end{IEEEkeywords}

\section{Introduction}


Modern processors commonly incorporate specialized circuitry to exploit data-level parallelism for accelerating compute-intensive operations. At the instruction set architecture (ISA) level, this is manifested as various SIMD or vector extension instruction sets, such as the SSE and AVX/AVX2/AVX-512\cite{intelIntelAdvanced} instruction sets for the Intel x86 architecture, the Neon\cite{ArmNeon} and SVE\cite{armsve} instruction sets for the Arm architecture, and the Vector extension for the emerging RISC-V architecture\cite{riscv}. To leverage the acceleration capabilities offered by this dedicated hardware, programs must employ platform-specific assembly instructions. Given the considerable difficulty of writing assembly code directly. Many programmers implement explicit vectorized algorithms in high-level languages (usually C/C++) using SIMD/Vector intrinsics provided by modern compilers, subsequently relying on the compiler to generate the corresponding platform-specific assembly instructions. Numerous high-performance computing libraries, such as OpenCV\cite{opencv_library}, Libjpeg-Turbo\cite{Libjpeg-Turbo_2010}, OpenBLAS\cite{openblas}, PyTorch/Eigen\cite{PyTorch, eigenweb}, and NCNN\cite{ncnn}, have adopted this methodology to achieve performance optimizations across various platforms. This diversity in instruction sets, each with its distinct intrinsic functions, necessitates that these libraries include multiple intrinsic-based vectorized implementations for a single algorithm to guarantee performance across different platforms. Consequently, this introduces substantial challenges for development and maintenance, particularly when a new instruction set architecture aims to expand its software ecosystem by enabling or porting a large number of vectorized algorithms across high-performance libraries.

\begin{figure}[htbp]
\centerline{\includegraphics[width=0.8\linewidth]{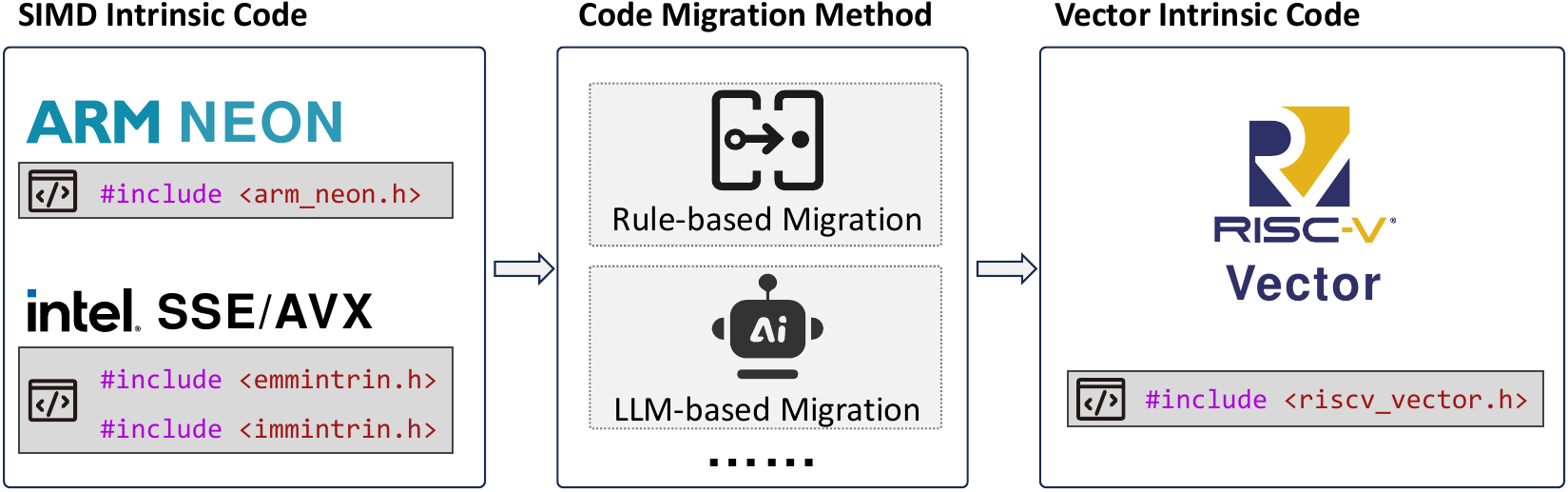}}
\caption{Example of Migrating Arm Neon Intrinsic Code to RVV}
\label{migration}
\end{figure}

Referring to intrinsic implementations from existing architectures in algorithm libraries is an effective way to reduce manual effort. Consequently, the direct migration of intrinsic code has become a prominent research direction. An example of a code migration task targeting the RVV extension is shown in Figure \ref{migration}. Although this field is still in its early stages of research, the feasibility of rule-based approaches has been explored by intrinsic migration tools such as sse2rvv\cite{sse2rvv} and neon2rvv\cite{neon2rvv}. However, due to the absence of a comprehensive benchmark that includes RVV intrinsics, the practical effectiveness of these tools has yet to be fully evaluated.

 LLMs have demonstrated exceptional capabilities in code generation. LLM-Vectorizor\cite{llm-vectorizer} and VecTrans\cite{vectrans} translate scalar C code into vectorized intrinsic code, showcasing the potential of LLMs in both generating and migrating intrinsic code. However, a benchmark for evaluating the quality of such migration remains absent. Existing vectorized benchmarks, such as ParVec\cite{ParVec} and LLaMeSIMD\cite{LLaMeSIMD}, do not include the emerging RVV intrinsics, making them unsuitable for assessing migration capabilities targeting the RISC-V architecture. More recently, He et al. proposed SimdBench\cite{simdbench}, the first benchmark for SIMD intrinsic code generation. It provides task descriptions, scalar C solutions, as well as correctness and performance tests, and presents a systematic evaluation of mainstream LLMs. However, this work focuses on generating intrinsic code from scalar C, not on migrating code between different intrinsic functions, and therefore contains no source intrinsic code for migration tasks.



To address this gap, we introduce VecIntrinBench, the first benchmark designed explicitly for intrinsic code migration to the RVV instruction set. It comprises 50 use cases sourced from high-performance libraries, authored by experts from the community, and covers intrinsic code from the RISC-V, Arm, and x86 architectures. Equipped with comprehensive correctness and performance tests, VecIntrinBench facilitates a practical evaluation of code migration quality for RVV intrinsics. We utilize our benchmark to evaluate both rule-based and LLM-based intrinsic code migration approaches. The results indicate that LLM-based methods outperform rule-based ones in terms of both correctness and performance. However, LLM-based approaches still suffer from limitations due to an incomplete understanding of RVV hardware characteristics and intrinsic specifications \cite{rvv}. These findings provide guidance for future progress in intrinsic code migration, thus contributing to the continued growth of the RISC-V software ecosystem. VecIntrinBench is open-sourced at \url {https://github.com/hanliutong/VecIntrinBench}.

\section{Benchmark Construction}

\begin{figure}[htbp]
\includegraphics[width=\linewidth]{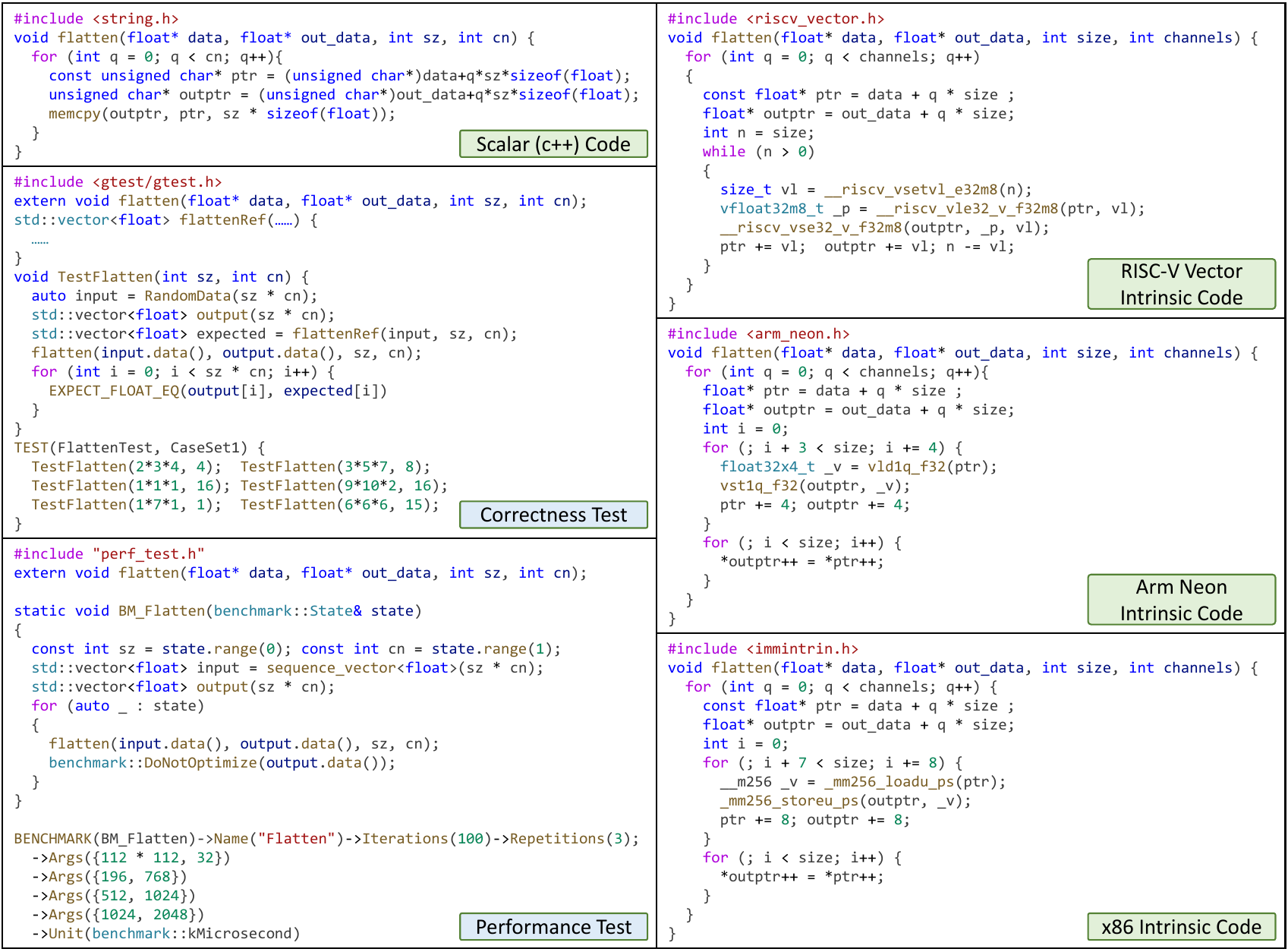}
\caption{Overview of a Case in VecIntrinBench}
\label{VecIntrinBench}
\end{figure}

VecIntrinBench is a benchmark designed to evaluate cross-architecture vector intrinsic migration. It is the first to incorporate RVV intrinsics, comprising 50 cases extracted from real-world applications. As illustrated by the sample case in Figure \ref{VecIntrinBench}, each use case represents the implementations of a specific algorithm and is accompanied by correctness and performance tests.

Regarding the code implementations, all 50 use cases include a scalar C++ version, an RVV intrinsic version, and an Arm Neon intrinsic version. A subset of these use cases is also implemented by x86 intrinsics. These use cases were collected from open-source repositories. To find repositories containing RVV intrinsic code, we searched for code with the keywords \texttt{\#include <riscv\_vector.h>} and \texttt{\_\_riscv\_v} through GitHub. From the search results, we identified functions within the NCNN, Libjpeg-Turbo, and OpenCV libraries that possessed implementations for multiple platforms, including RVV. These functions were incorporated into the benchmark, and we added correctness and performance tests for them. The source of each case and intrinsic implementations are detailed in Table \ref{Source}.

\begin{table}[]
\centering
\caption{The Cases Source in VecIntrinBench}
\label{Source}
\resizebox{\columnwidth}{!}{%
\begin{tabular}{cccccc}
\hline
\multirow{2}{*}{Repository} & \multirow{2}{*}{Number of cases} & \multirow{2}{*}{Scalar Code} & \multicolumn{3}{c}{Intrinsic Code} \\ \cline{4-6} 
              &    &            & RVV        & Arm Neon   & x86        \\ \hline
NCNN          & 22 & \checkmark & \checkmark & \checkmark & \checkmark \\
Libjpeg-Turbo & 17 & \checkmark & \checkmark & \checkmark &  -         \\
OpenCV        & 11 & \checkmark & \checkmark & \checkmark &  -         \\ \hline
\end{tabular}%
}
\end{table}

The scalar and intrinsic implementations for each case share a uniform function definition, enabling a straightforward switch between different versions by simply substituting the source file at compile time. 
This design is highly convenient for evaluating the correctness and performance across multiple implementations. Since all code is sourced from implementations by human experts within the open-source community, a high standard of code quality is generally assured. However, at the time of VecIntrinBench's construction, the RVV intrinsic code in the Libjpeg-Turbo repository was still undergoing review.


Correctness testing, driven by Google Test, verifies correctness by comparing the execution results of the generated code against expected values or the output of the scalar reference implementation. 
The correctness test cases generally consist of predefined edge cases and randomly generated inputs within valid ranges to ensure comprehensive evaluation of the function under test.
By linking the compiled object files of the function under test and the correctness test cases into an executable, the correctness of the generated function can be assessed. In general, the generated code is considered correct only when it can be compiled, executed, and passes all test cases successfully.

Performance testing is driven by Google Benchmark to facilitate the creation of input data at various scales for performance evaluation. To ensure the accuracy and reliability of the results, each performance test case is executed for 100 iterations, and this process is repeated 3 times, recording the average execution time of the function under test.


\begin{equation}
pass@k=\mathbb{E}_{problem} \left [ 1-\frac{\binom{n-c}{k} }{\binom{n}{k} }  \right ] 
\label{passk}
\end{equation}

The \texttt{pass@k} metric and speedup ratio are used to evaluate the effectiveness of code migration methods and the performance of the migrated artifacts, respectively. As defined by OpenAI and shown in Equation \ref{passk},  \texttt{pass@k}\cite{evaluatingLLMs} is a metric for evaluating the problem-solving ability of LLMs. In \texttt{k} generated code candidates, if at least one passes all test cases, the corresponding problem is considered passed. The \texttt{pass@k} score is the proportion of problems that are passed. Within VecIntrinBench, \texttt{pass@k} reflects the percentage of the 50 tasks that a migration method correctly handles in \texttt{k} attempts.


The speedup ratio is a conventional metric for assessing the performance of vectorized code. In VecIntrinBench, for each code that compiles and is validated for correctness, we measure its performance against a baseline: the execution time of the manually written native intrinsic code. The resulting speedup ratio is used to evaluate the optimization quality of the valid migrated code. As formulated in Equation \ref{eq-speedi}, for a migration task on case \texttt{i}, a speedup ratio approaching $1.0\times$ indicates that the performance of the migrated code is close to that of a human expert's implementation. A speedup of more than $1.0\times$ is possible because the code in open source code repositories is not always optimally best due to factors such as developer implementation capabilities. Advanced migration methods may outperform vectorized implementations by human experts.

\begin{equation}
speedup_{i} = \frac{Native Code Cost_i}{Migrated Code Cost_i}
\label{eq-speedi}
\end{equation}

\section{Evaluation}


To investigate the cross-architecture porting capabilities of different code migration methods to RVV intrinsics, our experiments on VecIntrinBench evaluated the migration from Arm Neon intrinsics to RVV intrinsics. For the rule-based code mapping method, we selected the open-source solution neon2rvv. For the LLM-based method, we chose six mainstream LLMs; their names, tags, and reasoning capabilities are presented in Table \ref{model}. We set the temperature to 0.2, a value suitable for code-related tasks, and used default settings for all other parameters. For each case in VecIntrinBench, we provided its Arm Neon implementation as input to the LLMs and used a prompt to instruct them to migrate it to RVV intrinsics.

The migrated code was compiled using the GNU toolchain for RISC-V with the \texttt{-O3} optimization option. For cases that compiled successfully, we evaluated their correctness using the QEMU emulator with 128 and 256-bit RVV vector register length(VLEN) to ensure the generation of valid RVV intrinsic code. For the passed cases, we conducted performance evaluations on a RISC-V device that supports the vector extension, the SpacemiT MUSE Pi, with the VLEN of 256-bit.

\begin{table}
\centering
\caption{Evaluated LLMs on our VecIntrinBench.}
\label{model}
\resizebox{\columnwidth}{!}{%
\begin{tabular}{llll}
\hline
\textbf{Model}    & \textbf{Tag} & \textbf{Org.} & \textbf{Reasoning} \\ \hline
Claude-sonnet-4.5 & 20250929     & Anthropic     & Enable/Disable     \\
Grok 4            & 0709         & xAI           & Enable             \\
Gemini 2.5 Pro    & -            & Google        & Enable             \\
GPT-5 Codex       & -            & OpenAI        & Enable             \\
DeepSeek-V3.2     & Exp          & DeepSeek      & Enable/Disable     \\
Qwen3 Coder Plus  & -            & Alibaba       & Disable            \\ \hline
\end{tabular}%
}
\end{table}


\begin{figure}[htbp]
\centerline{\includegraphics[width=0.9\linewidth]{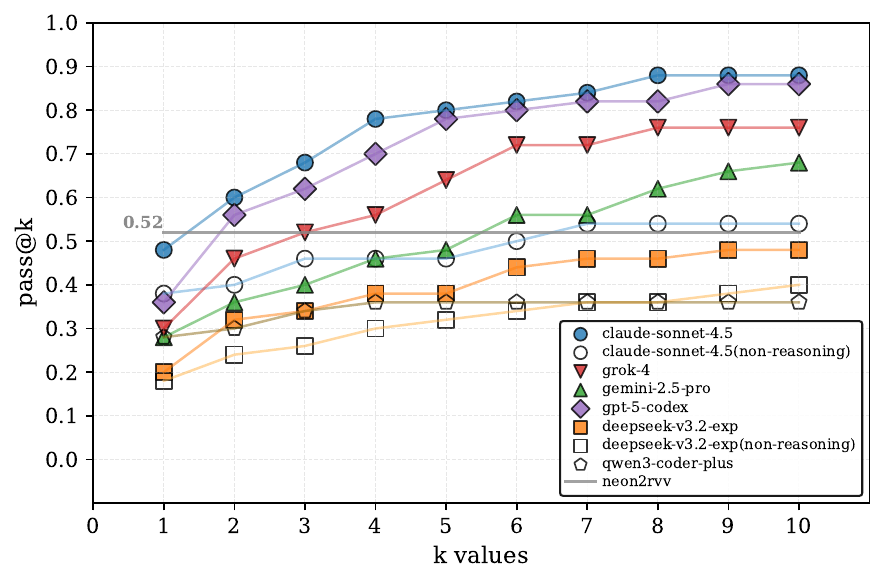}}
\caption{Effectiveness of code migration methods at different levels of k.}
\label{pass}
\end{figure}

In Figure \ref{pass}, we plot the  \texttt{pass@k} values (y-axis) against different values of \texttt{k} (x-axis) for the LLM-based migration methods. The results show that as \texttt{k} increases from 1 to 8, the  \texttt{pass@k} metric exhibits a significant upward trend, which indicates that today's advanced LLMs are generally capable of finding at least one viable migration solution within a limited number of attempts. When \texttt{k} is greater than 8,  \texttt{pass@k} tends to stabilize, which means that more code migrations do not help much in improving the pass rate of code migration. Different LLMs can correctly migrate 36\% to 88\% of the cases in VecIntrinBench with prompts, and the \texttt{pass@k} of the reasoning model has a significant advantage. In contrast, the rule-based neon2rvv library successfully migrated 52\% of the tasks (26/50) in VecIntrinBench.


\begin{table}
\centering
\caption{Migration results of the LLM-based method.}
\label{results}
\resizebox{\columnwidth}{!}{%
\begin{tabular}{lll}
\hline
\textbf{Migration Results}               & \textbf{Count} & \textbf{Proportion} \\ \hline
Passed                                   & 1311           & 32.78\%             \\
Compilation Failed: Undeclared functions & 1542           & 38.55\%             \\
Compilation Failed: Argument mismatch    & 503            & 12.58\%             \\
Test Failed                              & 326            & 8.15\%              \\
Compilation Failed: Others               & 175            & 4.38\%              \\
Compilation Failed: Incomplete code      & 129            & 3.23\%              \\
No RVV Intrinsic                         & 14             & 0.35\%              \\ \hline
\end{tabular}%
}
\end{table}

\begin{figure*}[htbp]
\centerline{\includegraphics[width=0.9\linewidth]{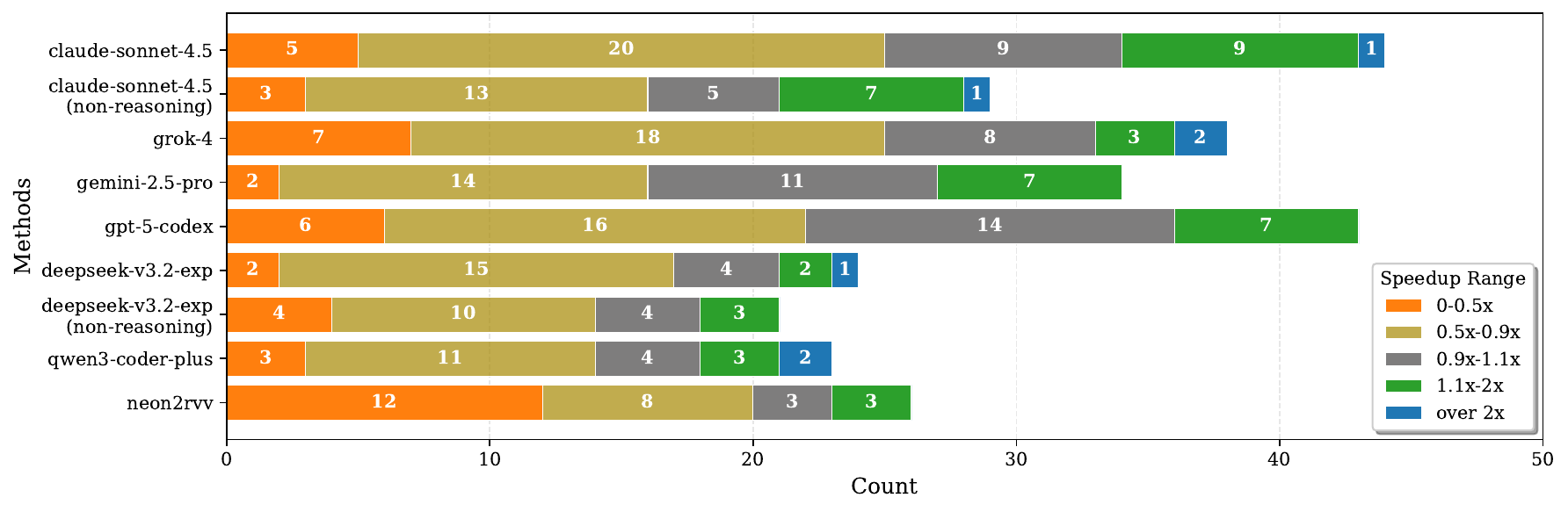}}
\caption{Performance Speedup Distribution for Passed Vector Intrinsics.}
\label{speedup_distribution}
\end{figure*}

A detailed examination of these cases, with the causes of failure categorized in Table \ref{results}, reveals that the most significant reason for failure is the incorrect use of RVV intrinsic functions. This includes generating functions that do not exist in the RVV Intrinsic specification and using incorrect function parameter types. These errors reflect that the LLMs' understanding of the RVV intrinsic API specification is not sufficiently accurate and that they exhibit systematic biases, particularly in areas involving vector types and their associated data widths. Some tasks also failed due to not passing the correctness tests or for other reasons. This indicates that maintaining semantic equivalence during the code migration process remains a challenge for LLMs, especially when dealing with features that have significant architectural differences, such as vector length agnosticism and rounding modes.


We further investigated the performance differences between the migrated code and the code written by human experts. Overall, the LLM-based migration methods demonstrated superior performance compared to the rule-based approach.
As shown in the distribution of speedup ratios in Figure \ref{speedup_distribution}, LLM-based methods produce better results. In the figure, the colors represent performance outcomes: \textcolor[HTML]{ff7f0e}{Orange} indicates that the performance of the migrated code is significantly worse than the native code. \textcolor[HTML]{a88b06}{Yellow} denotes a speedup ratio between 0.5 and 0.9, suggesting that the migrated code has room for optimization. This may occur because the RVV intrinsic code was migrated from a vectorized implementation for another architecture and thus fails to fully utilize RVV's specific hardware features and vector register resources. \textcolor[HTML]{2ca02c}{Green} and \textcolor[HTML]{1f77b4}{Blue} indicate that the migrated code outperforms the native code, likely due to the suboptimal quality of the native code. \textcolor[HTML]{7E7D7D}{Gray} represents cases where the performance of the migrated code is nearly identical to that of the native code. Compared to the rule-based method (neon2rvv), LLM-based methods resulted in fewer cases falling into the severely degraded performance category (\textcolor[HTML]{ff7f0e}{orange}) and more cases achieving similar or better performance (\textcolor[HTML]{7E7D7D}{gray}, \textcolor[HTML]{2ca02c}{green} and \textcolor[HTML]{1f77b4}{blue}).

\section{Discussion and Conclusion}




By comparing LLM-based and rule-based approaches, we observe that rule-based code mapping can partially accomplish cross-architecture intrinsic code migration targeting RVV.
However, they still require expert developers to further bridge the architectural differences in instruction mapping rules to enhance usability and performance. 
LLM-based approaches show great potential for intrinsic code migration, as current advanced LLMs have already outperformed rule-based methods.
However, they are more costly to use than rule-based methods and also face the following challenges:(1) The calls to the target intrinsic may deviate from the formal specification. Since modern compilers can provide modification suggestions, building a code migration agent with an error feedback mechanism will help solve this problem. (2) A lack of deep understanding of the semantic differences in the underlying hardware. To this end, future improvements should focus on enhancing the models' ability to learn domain-specific knowledge.

We believe that future work in the following directions would be beneficial for advancing intrinsic code migration:
(1) Introducing feedback mechanisms and Retrieval-Augmented Generation (RAG)\cite{rag} into the code migration process to provide additional information that helps LLMs reduce incorrect RVV intrinsic calls.
(2) Developing training datasets that include intrinsics from RVV and other architectures. This data can be used to fine-tune LLMs, enabling them to better adapt to RVV code migration.
(3) Exploring the fusion of LLM-based and rule-based methods. Intrinsic mapping rules can provide a strong reference for LLMs, thereby addressing the primary issue of incorrect function calls. In turn, the code generation capabilities of LLMs can compensate for the insufficient coverage of manually written mapping rules and improve performance issues caused by a lack of optimization.

In summary, we proposed VecIntrinBench, the first open-source intrinsic code benchmark including the RVV. It comprises 50 code migration tasks. We conducted code migration experiments on VecIntrinBench to evaluate the correctness and performance of both rule-based and LLM-based methods. The evaluation results demonstrate that LLM-based methods using advanced reasoning models achieve a migration pass rate ~20\% higher than the rule-based method with no more than 10 attempts, and the performance is significantly better. Our analysis identifies the key challenges in the current code migration landscape and outlines several promising directions for future advancement.

\bibliographystyle{IEEEtran}
\bibliography{ref}

\end{document}